\documentclass[twocolumn,aps,prl,showpacs]{revtex4}

\usepackage{graphicx}% Include figure files
\usepackage{subeqnarray,amsmath}% Number equations with letter
\usepackage{dcolumn}% Align table columns on decimal point
\usepackage{bm}% bold math
\usepackage{color}
\definecolor{pink}{rgb}{1,0.5,0.5}

\begin{document}

\preprint{APS/123-QED}

%\title{Potential profile in molecular conductors}
\title{Electrostatic potential profiles of molecular conductors}

\author{G. C. Liang}
\email{liangg@ecn.purdue.edu}
\author{A. W. Ghosh}
\email{ghosha@ecn.purdue.edu}
\author{M. Paulsson}
\author{S. Datta}
\affiliation{ School of Electrical and Computer Engineering, Purdue
University, W. Lafayette, IN 47907}%

\date{\today}% It is always \today, today,

\widetext
\begin{abstract}
The electrostatic potential across a short ballistic molecular conductor
depends sensitively on the geometry of its environment, and can affect
its conduction significantly by influencing its energy levels and wave
functions. We illustrate some of the issues involved by evaluating the
potential profiles for a conducting gold wire and an aromatic phenyl
dithiol molecule in various geometries. The potential profile is obtained
by solving Poisson's equation with boundary conditions set by the contact
electrochemical potentials and coupling the result self-consistently
with a nonequilibrium Green's function (NEGF) formulation of transport.
The overall shape of the potential profile (ramp vs. flat) depends on
the feasibility of transverse screening of electric fields. Accordingly,
the screening is better for a thick wire, a multiwalled nanotube or a
close-packed self-assembled monolayer (SAM), in comparison to a thin
wire, a singlewalled nanotube or an isolated molecular conductor. The
electrostatic potential further governs the alignment or misalignment
of intramolecular levels, which can strongly influence the molecular
I-V characteristic. An external gate voltage can modify the overall
potential profile, changing the current-voltage (I-V) characteristic from
a resonant conducting to a saturating one. The degree of saturation and
gate modulation depends on the availability of 
metal-induced-gap states (MIGS) and on
the electrostatic gate control parameter set by the ratio of the gate
oxide thickness to the channel length.  
\end{abstract} 
\bigskip

\pacs{PACS numbers: 85.65.+h, 73.23.-b,31.15.Ar}
%31.15.Ar Ab initio calculations
%81.07.Nb Molecular Nanostructures
%81.07.Lk Nanocontacts
%85.65.+h Molecular electronic devices
%72.10.Bg General formulation of transport theory
%72.20.Dp General theory, scattering mechanisms of conductivity
%73.23.-b Electronic transport in mesoscopic systems
%73.40.Sx Metal-semiconductor-metal structures
%73.63.-b Electronic transport in mesoscopic or nanoscale materials and structures
%2col
%end of wide text
\maketitle

Recently there has been considerable progress in the experimental
analysis \cite{rReed,rTiann,rReich,rMetzger} and theoretical modeling
\cite{rRatner,rDamle,rLangAvouris,rGuo,rPalacios} of molecular
electronic devices. Conduction through a molecule depends on its
intrinsic chemistry, as well as external influences such as the
electrode geometry, charging and bonding at the electrodes. In
particular, the electrostatic potential profile in molecular devices is
a quantity of great interest, because it contributes to the
self-consistent field in the molecular Hamiltonian, thereby influencing
the electronic properties and regulating the flow of current. 
The spatial variation of the electrostatic potential carries
nontrivial information about molecular screening, the presence of
impurities and Schottky barriers, as well as features related to the
alignment of energy levels within the molecule
\cite{rAviramRatner,rNDR}.  Furthermore, the long-ranged nature of
electrostatic forces allows the molecular levels and wavefunctions to
be tuned remotely, with a gate electrode for example.

It is often stated that the voltage drop in a ballistic conductor needs
no discussion, for all the drop must be localized at the interfaces
with the contacts. While this is true in some sense for the
{\it{electrochemical potential profile}}, $\mu(\vec{r})$, it is
certainly not true for the {\it{electrostatic potential}},
$\phi(\vec{r})$; in fact, the two profiles can differ widely in
low-dimensional materials with long screening lengths. We will try to
bring out this distinction later using a simple classical diffusion
model, since $\mu(\vec{r})$ is a tricky concept to define under more
general conditions of transport. The electrostatic potential
$\phi(\vec{r})$, by contrast, is a clearly defined concept even for
quantum transport far from equilibrium. This paper is essentially about
$\phi(\vec{r})$. Under certain conditions the electrostatic potential
profile can have significant effects on the current-voltage
characteristics of a conductor \cite{rTian}.  In a standard
two-terminal molecular conductance measurement \cite{rReed,rTiann}, for
example, the precise division of the applied bias between the source
and drain contacts can cause the I-V to change from symmetric to
asymmetric, with the conductance gap determined either by the molecular
levels exclusively or by the contact Fermi energy in addition
\cite{rparadigms}. Such
a sensitive dependence of current conduction on the potential profile
is not typical in mesoscopic physics, but quite routine in
molecular electronics. We will try to provide simple insights to describe
such conditions.

Direct measurement of the electrostatic potential profile of a molecular
conductor is challenging, given its small size. Attempts at direct
AFM-based or potentiometric measurement of the profile have been limited
to long ($\sim 0.3 - 3$ $\mu$m) carbon nanotubes \cite{rBockrath} or
organic molecular solids \cite{rfrisbie}, where the profile has been
determined mainly to be flat inside the molecular system, with the
voltage drop largely occurring at the contacts. Similar profiles have
been theoretically postulated or invoked in various semi-empirical
treatments of molecular conduction \cite{rRatner,rTian}. In sharp
contrast, a ramp-like potential across a molecular wire has been
calculated by semi-empirical \cite{rnitz}, as well as a number
of first-principles density functional theory (DFT)-based simulations
\cite{rDamle,rLangAvouris,rGuo}. It is further reasonable to expect that
potential variations on an atomic scale would be influenced by atomistic
features on the contacts and surrounding molecules.  The nature of the
potential profile thus needs to be sorted out.

\begin{figure}
\vspace{3.0in}
{\includegraphics{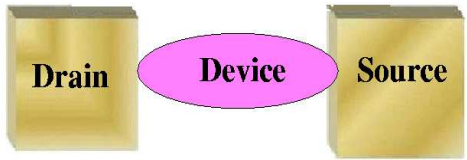}}
\vskip -1.8in
\caption{Schematic of two-terminal geometry, with metal
contacts of infinite cross section coupled to an active device. The
device will be replaced by a gold wire of various shapes throughout the
rest of the paper, and by a phenyl dithiol (PDT) molecule towards the
end. Part of the contact material may be included in the device in 
addition.}
\label{f0}
\end{figure}

In this paper we perform a fully quantum kinetic, atomistic treatment
of the electrostatic potential profile across a prototype molecular
conductor, and examine its influence on experimentally measurable
current-voltage (I-V) characteristics of the conductor. The prototype
conductor we investigate is either a gold atomic chain or a phenyl dithiol
molecule, attached to a cluster of metal atoms on both ends and sandwiched
between metal contacts of infinite cross section (Fig.~\ref{f0}). 
The ``trivial'' Laplace part $\phi_L$ of the potential profile
depends on the geometry and dielectric constants of the molecule and the
electrodes. This part is expected to dominate the electronic properties
for insulators and molecular conductors away from resonance, and can
be modulated externally with a gate, leading potentially to transistor
action. Furthermore, the Laplace part determines the alignment of energy
levels localized on different parts of the molecule, which could drive
interesting quantum effects such as current rectification or negative
differential resistance (NDR). In contrast, the quantum contributions
to the potential through the Poisson term $\phi_p(\rho)$ are affected by the
atomicity of the device. The Poisson part becomes important
in conducting systems such as metallic wires, metallic nanotubes and
molecules driven into resonance with a source-drain bias or turned on
with a gate.  $\phi_p(\rho)$ contains nontrivial physics, including
self-consistent charging, screening and Friedel oscillations.

Section I provides a semiclassical description of conduction that
outlines the difference between the electrochemical and the electrostatic
potentials.  Section II describes the nonequilibrium transport formalism
that we employ, and the model geometries and Hamiltonians that we use.
Section III calculates the voltage drop within the contacts, illustrating
that only a fraction of the applied bias drops across a charge neutral
region including the molecule and the contacts.  Section IV discusses
the dependence of the potential profile on the thickness of the wire,
specifically addressing the feasibility of adequate amount of transverse
screening.  Similar issues related to ineffective screening in 1-D have
been raised in the context of carbon nanotubes \cite{rTersoff-Odintsov}.
Section V addresses the importance of screening by the environment, such
as in a self-assembled monolayer (SAM) of molecular wires.  In section
VI we look at potential variations within the molecule, which although
small could nevertheless significantly influence observable electronic
properties, such as the current-voltage (I-V) characteristics. Finally
in Section VII, we talk about the influence of remote contacts such
as a gate electrode on the potential profile, which in turn determines
three-terminal transistor I-Vs. We summarize our results in section VIII.

\section{I. Semiclassical description of potential profile}

In this section, we present a simplified semiclassical, continuum
description of current conduction, invoking macroscopic parameters
such as the conductivity tensor and the dielectric constant. The aim
of this section is to provide an elementary
description of the distinction between
electrochemical and electrostatic potentials. For molecular systems
one cannot use such a macroscopic description, and a proper quantum
kinetic description needs to be invoked (section II).

A semiclassical description involves defining a conductivity
tensor $\sigma(\vec{r})$. The electrochemical potential satisfies the
{\it{equation of continuity}}, while the electrostatic potential satisfies
{\it{Poisson's equation}}:
\begin{subeqnarray}
\label{e000}
\slabel{e000a}
\vec{\nabla}\cdot\left(\sigma \vec{\nabla}\mu\right) &=& 0\\
\slabel{e000b}
\vec{\nabla}\cdot\left(\epsilon \vec{\nabla}\phi\right) &=& -e^2\left[n(\vec{r})-N_D\right]
\end{subeqnarray}
where 
$\epsilon$ is the dielectric
constant, $N_D$ is the dopant density, and 
$n(\vec{r})$ is the electron density:
\begin{eqnarray}
n(\vec{r}) &=& \int dED(E,\vec{r})f_0(E+\phi(\vec{r})-\mu(\vec{r})) \nonumber\\
f_0(E) &=& \left[1 + \exp{(E/k_BT)}\right]^{-1},
\label{rn}
\end{eqnarray}
$D(E,\vec{r})$ representing the local electron density of states (LDOS).
A conductivity mismatch at the device-contact interfaces, such as that
generated by a variation in the cross-sectional geometry or the doping
profile, allows us to hold the contact electrochemical potentials at fixed
voltages under bias, dropping $\mu$ almost entirely at the interfaces
for a ballistic device \cite{rDattabook}. This variation $\delta\mu$
influences the local charge distribution $\delta n(\vec{r})$ through the
chemical potential $\mu - \phi$ (Eq.~\ref{rn}), so that the electrochemical potential
profile $\mu(\vec{r})$ in effect acts as the driving force for the
electrostatic potential profile $\delta\phi$ in Eq.~\ref{e000b}. In
linear response, $\delta n(\vec{r}) \approx D_0\left(\delta\mu -
\delta\phi\right)$, where $D_0$ is the density of states (DOS) at the
Fermi energy.  Poisson's equation (\ref{e000b}) then reads:
\begin{equation}
\left(\nabla^2 - {{e^2D_0}\over{\epsilon}}\right)\delta \phi 
= - {{e^2D_0}\over{\epsilon}}\delta \mu
\end{equation}
indicating that $\delta\phi$ is given by a convolution of $\delta\mu$ and
a screening function that varies on a Debye (Thomas-Fermi) lengthscale
given by $\lambda_D \approx \sqrt{\epsilon/e^2D_0}$ \cite{rDebye}.
(An analogous expression can be invoked to describe screening by surface
states \cite{rSaslow}). 
{\it{The electrostatic potential thus has a slower spatial variation
than the electrochemical potential profile}} \cite{rMcLennan}.
In metallic conductors with a high DOS and correspondingly small Debye
length the two profiles track each other in order to avoid large charge
buildups. In contrast in a semiconductor or insulator having small DOS
inside the bandgap the Debye length can be quite large, so that the
two potentials can vary on widely different length scales \cite{rpn}.
A molecular conductor is intermediate between the two limits, acting as
an insulator when the contact electrochemical potentials are off-resonant
with the levels, and as a conductor on-resonance.  One can thus have
a nontrivial electrostatic potential profile even in a ballistic molecular
device where the electrochemical potential does not vary spatially at all,
except at the contact-molecular interfaces.

Having illustrated the basic distinction between the two kinds of
voltage-drop, we now move onto a rigorous quantum kinetic and atomistic
description of it.

\section{II: Quantum kinetic Formalism}

{\it{Basic equations}}.
Eq.~\ref{e000a} needs to be modified to solve for the full quantum
transport under bias. At equilibrium, one could still use the concept of
an electrochemical potential, defined by the contact Fermi energy $E_F$.
The electronic local density of states (LDOS) are obtained by solving
Schr\"odinger's equation for the molecular Hamiltonian, supplemented by
a self-consistent potential $U_{\rm{SCF}}$, and filling the eigenstates
according to equilibrium statistical mechanics: \begin{eqnarray} &&\left[H
+ U_{\rm{SCF}}\right]\Psi_\alpha(\vec{r}) = E_\alpha\Psi_\alpha(\vec{r})
\nonumber\\ &&n_{\rm{eq}}(\vec{r}) = \sum_\alpha n_\alpha(\vec{r}) =
\sum_\alpha |\Psi_\alpha(\vec{r})|^2f_0(E_\alpha-E_F).  \label{r0eq}
\end{eqnarray} In general, the self-consistent potential $U_{\rm{SCF}}$
consists of a Hartree-term obtained from the electrostatic potential
$\phi$ that is the solution to Poisson's equation \ref{e000b} with
$n$ replaced by $n_{\rm{eq}}$, and additional exchange-correlation
contributions that need to be incorporated through an appropriate
ab-initio technique. The expression for $n$ can be recast in the
form Eq.~\ref{rn} by recognizing that the LDOS $D(E,\vec{r}) = \sum_\alpha
|\Psi_\alpha(\vec{r})|^2\delta(E-E_\alpha)$.

Under bias, the contact electrochemical potentials separate,
and the system is driven out of equilibrium in its bid to equilibrate
with both contacts, causing a current flow. Under these nonequilibrium
conditions an unambiguous common electrochemical potential is hard to
define. One could, however, describe transport in terms of {\it{groups}}
of electrons that are separately in equilibrium with the two contacts
\cite{rDattabook}.
Furthermore, the contacts could fill the energy levels in a correlated
way, so that one needs to deal with the full nonequilibrium density
matrix $\rho(\vec{r},\vec{r}^\prime)$ of which the electron densities
$n_\alpha(\vec{r})$ form just the diagonal parts. A formal way of
handling these issues is by using the nonequilibrium Green's function
(NEGF) prescription \cite{rDattabook}.

In the NEGF formalism, one deals with retarded, advanced, lesser and
greater single-particle Green's functions \cite{rDattabook}. 
The retarded Green's function matrix is
\begin{equation}
G(E) = \left[ES - H - U_{\rm{scf}}(\rho) - \Sigma_S(E) - \Sigma_D(E) \right]^{-1}.
\label{eGreen}
\end{equation}
where $H$ is the single-particle molecular Hamiltonian in an
appropriate basis set, $S$ is the overlap matrix in that basis-set, the
self-consistent potential $U_{\rm{scf}}$ is dominated by the
electrostatic potential $\phi$ (and in principle, includes
exchange-correlation effects too), and the self-energy matrices
$\Sigma_{S,D}$ represent the influence of scattering by the source (S)
and drain (D) contacts. The contact self-energies can be calculated
for electrodes of given geometry and surface bonding \cite{rDamle},
yielding level broadenings $\Gamma_{S,D}$:
\begin{equation}
\Gamma^{}_{S,D} = i\left[\Sigma^{}_{S,D} - \Sigma_{S,D}^\dagger\right].
\label{ebroad}
\end{equation}
The nonequilibrium density matrix $\rho$ determining $\phi(\rho)$ is 
calculated self-consistently within the NEGF formalism using
the lesser Green's function as \cite{rDattabook,rDamle}:
\begin{equation}
\rho = \int dE\left[-iG^<(E)/2\pi\right],
\label{eGreen2}
\end{equation}
with the electron density representing its diagonal component (see
Eq.~\ref{eCharge}).
The function $-iG^<(E)$ describes how the molecular states are
filled in a correlated way by the two contacts, and is itself
determined by the retarded Green's function 
$G$ (Eq.~\ref{eGreen}), the broadenings $\Gamma_{S,D}$, and the
contact electrochemical potentials $\mu_{S,D}$:
\begin{eqnarray}
-iG^< &=& G\left[f_S\Gamma_S + f_D\Gamma_D\right]G^\dagger \nonumber\\
f_{S,D}(E) &=& \Biggl(1 + \exp{\Biggl[{{E-\mu_{S,D}}\over{k_BT}}\Biggr]}\Biggr)^{-1}.
\label{eGreen5}
\end{eqnarray}
Since we {\it{explicitly}} include any asymmetry in electrostatic
coupling with the contacts through $\phi(\rho)$ and in quantum coupling
through $\Gamma_{S,D}$, we dispose $\mu_{S,D}$ symmetrically about the
equilibrium contact Fermi energy $E_F$ under a source-drain bias $V$:
$\mu_{S,D} = E_F \mp eV/2$. {\it{The contact electrochemical potentials help fix
the boundary condition (Dirichlet or Neumann) for Poisson's equation in
the contacts, so that the electrostatic potential $\phi(\rho)$ within
the device can now develop on its own self-consistently under the applied
bias}} \cite{rGauge}.

The NEGF density matrix equations (Eqs.~\ref{eGreen2},\ref{eGreen5})
provide the nonequilibrium generalization of Eq.~\ref{r0eq}. The 
correspondence is evident by rewriting Eq.~\ref{r0eq} as 
\begin{eqnarray}
n_{\rm{eq}}(\vec{r}) &=& \rho_{\rm{eq}}(\vec{r},\vec{r})\nonumber\\
\left[\rho_{\rm{eq}}\right] &=& \int dEf_0(E-E_F)D(E) \nonumber\\
&=& \int {{dE}\over{2\pi}}f_0G(\Gamma_S+\Gamma_D)G^\dagger
\label{erq}
\end{eqnarray}
where $[\rho]$ and $D$ are the charge density matrix and the density of
states matrix whose diagonal entries give the LDOS and the trace gives
the total DOS.  The self-consistently converged Green's function $G$ is
then used to obtain the current.  For coherent transport the NEGF current
expression formally resembles that in Landauer theory \cite{rDattabook}:
\begin{eqnarray}
I &=& {{2e}\over{h}} \int_{-\infty}^{\infty} dE~T(E)
\left[f_S(E) - f_D(E)\right] \nonumber\\
T &=& {\rm{Trace}}\left(\Gamma_SG\Gamma_DG^\dagger\right).
\label{eGreen3}
\end{eqnarray}
Given an appropriate Hamiltonian $H$, self-consistent
potential $\phi$ and self-energies $\Sigma$,
eqns.~\ref{eGreen},\ref{ebroad},\ref{eGreen2},\ref{eGreen5},\ref{eGreen3}
yield a complete set of equations allowing us to obtain the
nonequilibrium charge and current densities under bias. Although the
equations describe coherent transport, NEGF allows us to handle incoherent
processes through additional self-energy matrices determined by the
corresponding scattering potentials. 

{\it{Incoherent processes}}.  Incoherent processes such as hopping
or phonon scattering can be included rigorously through a self-energy
in the NEGF prescription \cite{rLake}. A simple, phenomenological
way of including this would be to model each scattering center as a
floating voltage probe with which the electron equilibrates locally,
motivated by B\"uttiker \cite{rButt,rDattabook}. Such a probe does not draw
any net current, but randomizes the phase of the incoming electrons by
reinjecting them into the device. The probe can be parametrized by two
quantities, its electrochemical potential $\mu_p$ and the self-energy
$\Sigma_p$. The Green's function expression in Eq.~\ref{eGreen}
includes the additional self-energy contribution $\Sigma_p$, while
the transmissions $T_{ij}$ between any of the three contacts is given
by $T_{ij} = {\rm{Trace}}\left(\Gamma_i G \Gamma_j G^\dagger\right)$,
$i,j=S,D,p$.  

Besides modifying the transmission, incoherent scattering also affects
the local electron density through $\mu_p$, which is fixed by requiring
that the net current drawn by the voltage probe is zero.  One could
model {\it{phase-breaking without energy relaxation}}, where the probe
Fermi functions $f_p(E)$ are adjusted at {\it{each energy}} such that
the current $I_p(E)$ drawn by the probe at every individual energy
channel is zero. The expression for the net transmission in this case
becomes relatively straightforward:
\begin{equation}
T(E) = T_{SD}(E) + {{T_{Sp}(E)T_{pD}(E)}\over{T_{Sp}(E) + T_{pD}(E)}}
\label{eButt}
\end{equation}
Alternately one could include {\it{energy-relaxation}}, adjusting the overall
probe electrochemical potential $\mu_p$ such that the net probe current
$I_p = \int dE I_p(E)$ obtained by integrating the current contributions
over various energy channels adds up to zero. One will then need to
modify Eq.~\ref{eGreen3} to include the contributions from the probe to the
net source-drain current. The probe self-energy $\Sigma_p(E)$ can be
calculated using various models for scattering relaxation.  For phonons
for example, we use the self-consistent Born approximation \cite{rTiann},
whereby the phonon self-energy depends recursively on the electron Green's
function. In the limit of low-energy phonons, this
amounts to setting $\Sigma_p = \Delta_0G$, $\Delta_0$ being the coupling 
between the molecule and the phonon bath. In general, however, the
expressions are more complicated \cite{rDattabook}.

{\it{Model Hamiltonians for device and contacts.}}
The above formalism is general, and requires an appropriate Hamiltonian
$H$ describing the intrinsic molecular chemistry, an adequate
treatment of the geometry and bonding at the contact surfaces
described by self-energy matrices $\Sigma_{S,D}$, as
well as a suitable self-consistent potential $\phi$ describing the
electrostatics of the device. In the past, we calculated these matrices
by modifying a standard quantum chemical software (GAUSSIAN98),
self-consistently coupled with a NEGF-based transport formalism \cite{rDamle}.
Such a modification allowed us to get I-V characteristics and potential
profiles using a density functional (DFT) description of both the
molecule and the contacts.  While DFT codes allow `first principles'
treatment of electron-electron interactions and quantum correlations
with no adjustable parameters, they are tedious and time-consuming.
Furthermore incorporating the Poisson boundary conditions typically
requires a large chunk of the contacts included along with the molecule
within our device \cite{rDamle3}. Alternatively, one needs to perform
repeated transformations between an orthogonal real space basis where
Poisson boundary conditions are readily incorporated, and a
non-orthogonal orbital basis suitable for describing the atomic
bondings and the molecular chemistry.

In order to bypass the complexities associated with implementing boundary
conditions within DFT, we will concentrate here on a much simpler
semi-empirical description of the device and contacts, described by
an orthogonal basis set $\{\Phi_\mu\}$ of one orbital per atom. Such a
basis could, for instance, describe the essential physics of molecular
transport through conjugated $p_z$ ($\pi$) electron systems, and $s$
electrons for metallic gold wires. In this paper, the scalar on-site and
hopping parameters are chosen to match experimental values of the Fermi
energy and density of states for gold, as well as the energy eigenvalues
and orbital shapes for phenyl dithiol (PDT) calculated using DFT within
the local density approximation (LDA) \cite{rDamle2} (parameters listed
in Table {\ref{table1}}). The source and
drain contacts are modeled as Au(111) surfaces, with the wire bonded
symmetrically to a surface triangle of gold atoms. The contact self
energies $\Sigma_{S,D}$ are obtained rigorously using a real space
recursive formalism described elsewhere \cite{rManoj,rDamle,rNDR}.

\begin{table}
\begin{tabular}{ccccc}
\colrule
&~On-site &~~ Hopping & ~~Hubbard &~~Bond-length \\ 
& (eV) & (eV) & (eV) & ($\AA$)\\
\colrule
Au-Au& -4.3 & ~8.75 & 11.13 & ~~2.885 \\
C-C & ~-4.42 &2.5 & 11.13 & ~~~1.4625\\
S   & ~ -6.32 & &~~9.94 & \\
S-C & & 1.5 & &1.8 \\
S-Au &&1.6 & &~~ 2.885\\
\colrule
\end{tabular}
\caption{Semi-empirical parameters used to simulate molecular wires.
The S-Au coupling is reduced from 1.6 to 0.8 in the
last section to simulate gating of a weakly contacted molecular wire.}
\label{table1}
\end{table}

{\it{Self-consistent potential.}}
The self-consistent potential $\phi(\rho)$ in Eq.~\ref{eGreen} can be
obtained for a given charge density matrix $\rho$ in a variety of ways.
We will ignore the evolution of the exchange-correlation potential
under bias (the equilibrium result is incorporated in $H$), and
concentrate only on the Hartree part \cite{rnote}, which depends only
on the diagonal elements $n$ of the density matrix.  A computationally
tedious but accurate way to obtain the Hartree contribution is the
direct solution of 3-D Poisson's equation numerically on a real space
grid, with appropriate Dirichlet/Neumann boundary conditions at the
boundaries (section IV). A faster way, that involves grid points only
on the charges and the contacts is the {\it{method of moments}} (MOM)
\cite{rmom}. To understand this, one starts by writing the solution to
Poisson's equation in matrix form in a suitable basis $\{\Phi\}$, and
then partitioning the system into the device (`d') and the contact
(`c') sectors:
\begin{eqnarray}
\left(\begin{array}{c} \phi_d \\ \phi_c \end{array}\right)  &=&
\left(\begin{array}{cc} U_{dd} & U_{dc}\\ U_{cd} & U_{cc} \end{array}\right) 
\left(\begin{array}{c} n_d \\ n_c \end{array}\right) \nonumber\\
n(\vec{r}) &=& \rho(\vec{r},\vec{r}) = \sum_{\mu\nu}\Phi^*_\mu(\vec{r})
\rho^{}_{\mu\nu}\Phi^{}_\nu(\vec{r}),
\label{eCharge}
\end{eqnarray}
$U$ representing the Hartree term \cite{rP}. Eliminating the contact
charge density $n_c$, one gets the device electrostatic potential that
is used in Eq.~\ref{eGreen} by solving the matrix equation:
\begin{eqnarray}
\phi(\rho) \equiv \phi_d &=&  U^{}_{dc}U^{-1}_{cc}\phi^{}_c +  \left(U^{}_{dd} - U^{}_{dc}U^{-1}_{cc}U^{}_{cd}\right)n_d^{} \nonumber\\
&=& \phi_L + \phi_P(\rho),
\label{eCharge2}
\end{eqnarray}
where the Laplace part is the solution to the homogeneous Poisson's
equation corresponding to {\it{zero charge in the device}} ($n_d =
0$) and specified contact potentials, while the inhomogeneous Poisson
part corresponds to the solution for given contact and device charges
and with {\it{zero potential on all the contacts}} ($\phi_c = 0$).
The advantage of MOM is that we need grid points only on the contacts
with specified potentials $\phi_c$ and on the device with specified
free charges $n_d$. The formalism can easily be extended to include
polarization charges in dielectric materials.

 In this article, the Coulomb matrices $U$
are obtained using the Pariser-Parr-Pople (PPP) model within
the Matago-Nishimoto approximation \cite{rMatago}:
\begin{equation} 
U(\vec{r}_i,\vec{r}_j) = {{e^2}\over{\left|\vec{r}_i -
\vec{r}_j\right| + 1/\gamma_{ij}}}
\label{eCharge3} 
\end{equation}
where $\gamma_{ij} = 0.5(U_{ii} + U_{jj})/e^2$.
The above term interpolates between asymptotic $1/r$ Coulomb repulsion at
large distances and on-site Hubbard repulsion $U_{ii}$ at short distances
(table \ref{table1})
(Alternate ways of handling the Hubbard term exist;
see for e.g. Mc. Lennan {\it{et al.}}, \cite{rMcLennan}, Appendix).
The effect of image charges on the contact surface atoms is already
included in Eq.~\ref{eCharge2} through the second term in brackets in
the expression for $\phi_p(\rho)$. Additional, possibly negligible,
image contributions from deeper within the contacts are added in by hand
by generating a series of contributions similar to Eq.~\ref{eCharge3}
due to electrostatically infinite planar contacts \cite{rpauls}.

The boundary conditions on Poisson's equation are set by the
electrochemical potentials $\mu_{S,D}$. One can either use Neumann
boundary conditions, requiring overall charge neutrality deep inside
the contacts (section IV).  Alternatively, one could enforce Dirichlet
boundary conditions, setting the electrostatic potentials equal to the
electrochemical potentials deep inside the contact (rest of the paper).

Having laid out our general framework and the description of our model
Hamiltonians, we will first try to determine how much of the applied 
bias across a metal-molecule-metal heterostructure drops within the
contacts themselves.

\section{III. How much voltage drops across the molecule?}

\begin{figure*}
\vspace{3.7in}
{\includegraphics{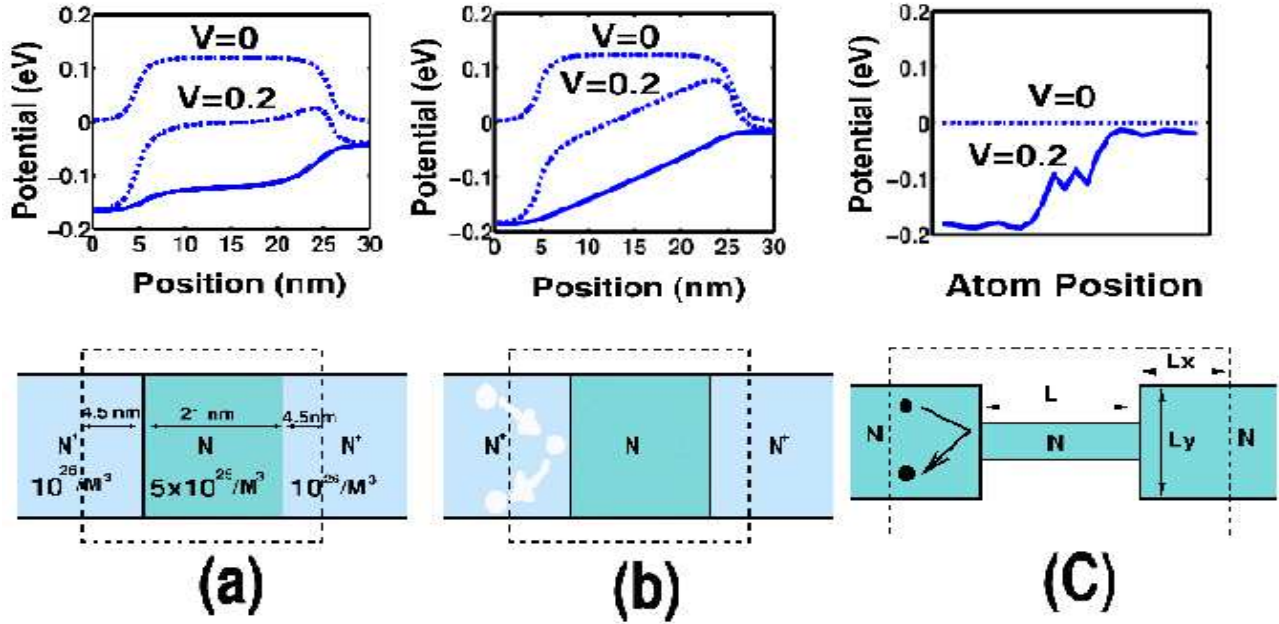}}
\hskip -2.5in
\caption{Electrostatic potential profile in various
structures under zero-field boundary conditions (ZFB). Part of the
contacts (blue) has been incorporated into the device region within
dashed lines. (a) Charge
neutrality restrictions due to ZFB in a silicon (n$^+$nn$^+$) structure
causes the source electrostatic and electrochemical potentials under a
0.2 V bias to separate, thus dropping some of the applied voltage in
the contacts.  The voltage ``loss'' can be reduced by (b) increasing
scattering or by (c) a geometric dilution (see text), which reduces
this mismatch. The structures in (a) and (b) have the same geometry and
doping properties, shown in the figure. Structure (c) is simulated by
using gold parameters (Table 1). The channel consists of a four atom gold
chain, while each contact cluster within the device region consists of
a $10 (L_x) \times 11 (L_y)$ simple cubic gold lattice.}
\label{fcontpot} 
\end{figure*}

The transport formalism described in section II applies equally to the
device and the contacts. To distinguish between the two, one needs a
dilution of modes in going from the contact to the device, such as a
drastic change in doping profile (Figs.~\ref{fcontpot}a,b) or a geometrical
narrowing of the cross-section (Fig.~\ref{fcontpot}c).  A slight
deviation from equilibrium inside the contact, integrated over its
multiple modes, is then sufficient to drive current through the device,
while keeping the contact electrochemical potentials essentially constant.
In terms of Eq.~\ref{e000a}, one needs a large conductivity inside the
contacts relative to the molecule in order to hold the electrochemical
potentials constant inside each contact and drop the applied voltage
bias across the molecule itself. For a ballistic device, however, part
of the applied voltage drops within the contacts, as we now describe.

To simulate the potential profile under bias across the device
as well as the contacts, we will consider an extended device that
actually incorporates a large enough part of the contacts to keep
it essentially charge neutral. Such an inclusion would automatically
take care of image charges for metal contacts or depletion/inversion charges
in semiconducting contacts, for example. However, charge neutrality
ensures that only a {\it{fraction}} of the applied bias appears across
the ends of the device, especially if it is ballistic, the rest dropping
deeper inside the contacts. It is hard to impose Dirichlet boundary
conditions on the electrostatic potential at the two ends without knowing
this ``lost'' fraction a-priori; a better way seems to be to {\it{just
fix the electrochemical potentials $\mu_{S,D}$ at the two ends, impose
Neumann-type zero-field boundary conditions (ZFB) on the electrostatic
potential consistent with Gauss' law and charge neutrality, and let
the potential profile develop on its own self-consistently}}.

To illustrate the voltage loss within the contacts described above,
we simulate  a n-doped device (Fig.~\ref{fcontpot}(a)) consisting of
a cluster of silicon atoms of infinite cross-section attached
to a n++ silicon contact of identical structure and cross-section
but with a heavier doping density. Part of the contact has been
incorporated into the extended device simulation region (dashed lines)
to account
for surface charges. The equilibrium ($V=0$) potential profile shows a
barrier on the device part that reflects the incoming electron waves from
the contacts. We determine the potential profile under bias by explicitly
solving 1-D Poisson's equation self-consistently on a real space grid
using the Newton-Raphson method \cite{rNR} with ZFB at the ends. Note also
that in the transport equations \ref{eGreen5},\ref{erq},\ref{eGreen3}
one needs to replace each Fermi function $f(E)$ by the corresponding
2-D Fermi function $F_{\rm{2D}}(E)$ obtained by summing the transverse
modes over the two irrelevant dimensions \cite{raa},\cite{rab}.  As is
evident from Fig.~\ref{fcontpot}a, applying a 0.2 V source-drain bias
lowers the drain electrochemical potential, which in turn lowers the
electrostatic potential barrier and leads to current flow. The net
self-consistent potential, $\phi(V=0.2) - \phi(V=0)$ across the wire is
shown in solid line, and exhibits significant screening due to charge
rearrangement within the n-doped region. Significantly, lowering the
drain electrochemical potential also lowers the source electrostatic
potential, implying that {\it{only a fraction of the applied 0.2 V
bias actually appears across the simulated region}}. To understand this,
we recall that in a ballistic device the current carrying electrons can be
grouped into $+k$ and $-k$ states that originate in the drain and source
contacts respectively \cite{rDattabook}. A bias changes the occupancies
of these states, so that a current flows due to their imbalance. Inside
the source contact too, a bias produces a similar imbalance between
$+k$ and $-k$ states, with the result that part of the electron states
inside the source end up being depopulated by the drain electrode. The
depopulation forces part of the applied bias to drop inside the source
contact in order to keep it charge neutral \cite{rDattaSL}. Consequently,
only a fraction of the applied bias occurs across the two ends.

The depopulation of the source contact is particularly noticeable for
ballistic transport, where the drain end can effectively empty out the
source end and lower its electrostatic potential significantly. This
observation seems to be in keeping with Landauer's idea that only a
fraction $R$ ($R$: device reflection coefficient) of the applied bias
actually appears across the active device, which in conjunction with
a current proportional to the transmission $T$ gives the celebrated
Landauer conductance for a single-moded structure minus the interfaces,
$G = (2e^2/h)(T/R)$ \cite{rlb}. It seems evident therefore that the way to eliminate
the voltage loss for a ballistic device is to increase the average
scattering ($R \rightarrow 1$), which would in effect isolate the source
and drain and prevent the latter from depopulating the former. This could
be achieved by increasing the doping density in the n$^{++}$ region of
Fig.~\ref{fcontpot}(a), by increasing scattering throughout the device
(Fig.~\ref{fcontpot}(b)), or by enforcing a geometric dilution with
a flared-out contact geometry (Fig.~\ref{fcontpot}(c)). In (b) we use
B\"uttiker probes to simulate scattering, while in (c) we solve the 2-D
Poisson equation to take care of the varying transverse cross-sectional
geometry, using the relevant $F_{\rm{1D}}$ function obtained by summing
the Fermi-Dirac function over the single irrelevant dimension. As we see
in both cases, a larger fraction of the applied bias appears across the
ends with a smaller voltage loss in the contacts, as evidenced by
the closer agreement between the electrostatic potential values at $V=0$
and $V=0.2$ V deep within the source \cite{rFCC}. Scattering in the contacts can
thus have a significant influence on the device conductivity, tending
to make it robust with respect to spatial variations in the interfacial
geometry \cite{rVenugop}.

In the simulations we present hereafter, we will  explicitly incorporate a
built-up, atomistic part of the contacts along with the molecule within
our device, enforcing thereby a geometric dilution. Accordingly, we
will assume that most of the applied bias appears across the ends, with
minimum voltage loss in the contacts \cite{rcontregions}. This
assumption will allow us to switch to Dirichlet boundary conditions
imposed on the potential profile, which ends up being computationally
easier to handle. Since the contact potentials are assumed specified
hereafter, we will use the MOM instead of explicitly solving Poisson's
equation on a real space grid.

\section{IV. Potential screening inside a wire: thick vs. thin}

We will now concentrate on the potential profile inside the wire,
and discuss how its overall shape is determined by broad, macroscopic
geometrical features such as the wire thickness and screening length. We
will see that screening is ineffective in a wire that is thinner than
its Debye length.  Later on, we will talk about finer details, determined
by the internal molecular structure of the wire itself.

\begin{figure}[ht]
\vspace{3.4in}
{\includegraphics{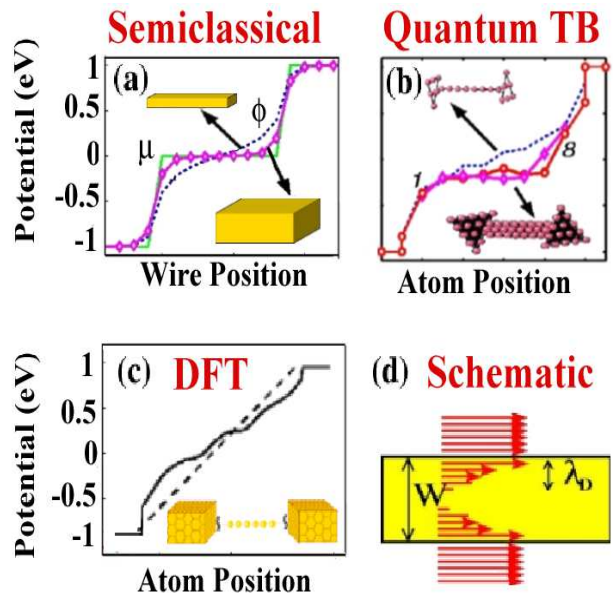}}
%\hskip -2in
%\vskip -1in
\caption{Electrostatic potential profile under two volts
applied bias for ballistic thin and thick wires (devices simulated
shown as insets) using (a) semiclassical modeling with a continuum
structure; (b) quantum tight-binding modeling with an atomistic
structure, and (c) fully ab-initio DFT modeling (adapted with
permission from \cite{rDamle}). Symbols represent the positions of
the atoms along the device. The electrostatic potential drop along
the wire decreases from thin to thick wire due to increased feasibility
of increased transverse screening when the wire thickness $W$ exceeds
the Debye length $\lambda_D$, for which the field lines are shown
schematically in (d). In the quantum calculations (b) and (c), the
potential profile exhibits Friedel oscillations superposed on the
long-wavelength Debye screening.  Self-consistent charging under bias
leads to a voltage asymmetry (see explanation in text as well as
caption of Fig.~\ref{f4.5}), although the I-V remains symmetric with
respect to bias direction.}
\label{f4} 
\end{figure}

Fig.~\ref{f4}(a) shows the potential profiles for a wire with a 2 volt
applied bias, using a semiclassical, continuum approximation
(Eq.~\ref{e000}). The wire is 9 atoms long, while the contact block
is 8 atoms long (only 3 atoms plotted) and 5x5 atoms in cross-section. The solid
line shows the average electrochemical potential along the wire, which
is spatially unvarying along the ballistic wire and drops only at the
contact interfaces.  Using this electrochemical potential $\mu$ as a
driving term in Eq.~\ref{e000b} gives the electrostatic potential
profile $\phi$.  The Debye lengths of the contact and the wire regions
are each assumed to be half the interatomic separation.  For a thick
slab representing the wire (3x3 atoms in cross-section, geometry shown lower
right), the potential profile (diamonds) is well-screened by charge
rearrangement in the material of the wire described by the Poisson part
$\phi_p(\rho)$. As we progressively thin the wire, however, we reach a
point where the thickness of the wire is no longer much smaller than
the Debye length (1x1 atom in cross-section, inset top left), at which point
transverse charge screening become ineffective. This causes the field
lines to penetrate in the transverse direction, generating a large
ramp-like electrostatic potential profile along the wire (dashed line),
essentially exhibiting just the Laplace solution $\phi_L$. A continuum
analysis by Nitzan
{\it{et al.}} leads to a similar conclusion, showing the
dependence of the slope of the potential on the wire length, thickness
and screening parameter \cite{rnitz}.

\begin{figure}[ht]
\vspace{2.8in}
{\includegraphics{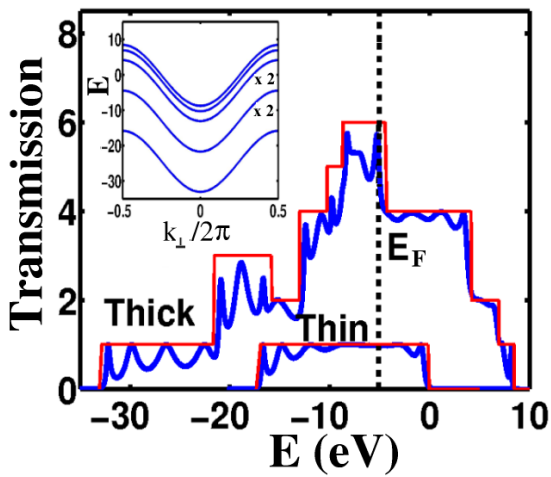}}
\caption{Equilibrium transmission of thin and thick wires
both show fine structure associated with Fabry-Perot end effects (blue
solid line) superposed on the overall structure associated with transverse
modes (mode transmissions in red). The multiplicity of channels for a
thick wire (subband structure shown in inset) arising due to quantization
of the transverse ($k_\perp$) quasimomentum leads to the overall structure
in its transmission (red). Under a small positive source bias, the drain
electrochemical potential adds less negative charge to the thick wire
than the source depletes, owing to the proximity of the Fermi energy
with a transmission mode bandedge that lies just above it in energy. The
net positive charge floats down the molecular levels relative to $E_F$,
giving rise to the asymmetry in the potential profiles in Fig.~\ref{f4}b.}
\label{f4.5} \end{figure}

In Fig.~\ref{f4}(b), we arrive at the same conclusions, albeit from a
more rigorous atomistic model Hamiltonian, including self-consisting
charging effects and the detailed bonding geometry at the Au(111)
contacts, and employing the full machinery of the quantum kinetic NEGF
transport formalism (Eqs.~\ref{eGreen}-\ref{eCharge2},\ref{eCharge3}).
The same two cases are studied as before, with a thin wire constructed
out of a chain of eight gold atoms connected to a triangle of FCC
Au(111) contact surface atoms (top left), and a thicker version
generated by enclosing the central wire in a sheath of six identical
gold wires around it (bottom right). The wires have quantum mechanical
couplings in the transverse direction as well, allowing charge flow in
that direction.  Contacting the wires to the 3-D gold leads leads to a
dimensionality mismatch and an associated work-function mismatch,
transferring about 0.8 electrons at equilibrium to the thin wire and 2
electrons to the thick wire (the charging energy per electron $U_0$ is
about 2-3 eV for the thin and 1.1 eV for the thick wire, so the work
function mismatch leads to a band adjustment of about 2 eV for both thin
and thick wires at equilibrium). The electrostatic potential profile along the
central wire progressively changes from flat (diamonds) to ramp (dashed
line) as the surrounding wires are removed \cite{rcontregions}.  A DFT
version with a six atom gold wire (Fig.~\ref{f4}c) \cite{rDamle} shows
a similar profile.  The screening along the transverse direction
(Fig.~\ref{f4}d) is feasible only if the wire thickness $W$ exceeds the
Debye length $\lambda_D$.  The oscillations in the potential profile
arise due to coherent Friedel oscillations (discussed later). A part of
the potential drops across the end triangle of contact atoms (symbols 1
and 8 referring to the end wire atoms in b). Interestingly, even the
surface potential along the thick wire (circles) shows screening
although it is exposed on least one side. We believe the other wires
screen the field over a large enough angle of the cross section that
the field penetrating from the exposed part is minimal.

%{\it{The important point to note is that a ramp-like electrostatic
%potential profile, as measured by an AFM tip \cite{rBockrath}, does
%not necessarily imply diffusive transport}}. Even a ballistic wire
%can give such a potential profile, provided the wire is thin enough
%(e.g. a ballistic single-walled carbon nanotube). An STM, however,
%probes the tunneling current rather than the electrostatic potential,
%so a ramp measured in $\mu$ implies scattering and diffusive transport
%in such cases.

\begin{figure}
\includegraphics[width=3.25in]{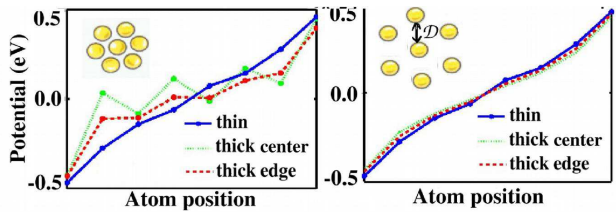}
%\vspace{2.0in}
%{\special{psfile=fig5.ps
%hscale=18.0 vscale=25.0}}
%\hskip -1.5in
%\vskip -1.5 in
\caption{Decrease in screening as a thick wire is pulled apart (same
geometry and applied bias as in Fig.~\ref{f4}b, with only the part
between atoms 1 and 8 shown here). For a closely packed coaxial wire,
the potential profile of the central wire was screened (Fig.~\ref{f4}b,
diamonds) compared to the thin wire (blue dotted line above, and blue
dashed line in Fig.~\ref{f4}b). The screening is reduced drastically
on pulling the wires just a little bit (left), since it eliminates
the quantum couplings among the wires and prohibits rearrangement
of screening charges in the transverse direction. The electrostatic
coupling between wires is longer ranged, and is eliminated once the
inter-wire separation ${\cal{D}}$ becomes comparable to the maximum
distance between a charge and the nearest source/drain ground plane,
i.e., half the wire length. The potential profile (right) of each wire
now resembles that of an individual thin wire.}
\label{f8}
\end{figure}

The potential profile for the thin wire is spatially symmetric, but
that of the thick wire shows a significant amount of asymmetry. This
asymmetry arises from self-consistent charging effects in the wire
associated with its quantum capacitance.  While the transmission of the
thin wire is effectively structureless and approximately unity over a
band (except for Fabry-Perot type resonances near the band-edges), the
transmission of the thicker wire has considerable structure owing to
the availability of various transverse modes that kick in and out at
various energies.  Fig.\ref{f4.5} shows the zero bias, non
self-consistent transmissions for a thin and a thick wire. The
transmissions show a rapid oscillatory structure (blue solid lines) due
to Fabry-Perot effects from the end superposed on a broader structure
that arises from the transverse modes of the wires (red solid lines).
For the thin 1-D wire, the transmission is unity between the band
edges.  For the thick wire, however, one can support multiple
transverse modes associated with the seven wires (subband structure
shown in inset; there are seven subbands, of which some are doubly
degenerate, denoted by `x 2').  At 2 V bias the existence of an edge in
the transmission mode spectrum above $E_F$ causes the source
electrochemical potential to add less charge to the wire than the drain
removes, so about 0.08 electrons are removed from the thick wire. The
average single-electron charging energy of the thick wire, $\sim$ 1.1
eV/electron causes a shift in average potential profile associated with
the charge depletion by $\sim 0.09$ eV.  The shift gets localized on
only the central 5 atoms or so where charge depletion is most, so the
local electrostatic potentials at those points get lowered by about
$0.14-0.2$ V relative to $E_F$, as in Fig.~\ref{f4}b, making the
overall electrostatic potential profile asymmetric. Such an asymmetry
due to charging arises whenever the transmission spectrum has
significant structure, as for a multimoded wire or a molecular
conductor like PDT with distinct asymmetric highest occupied (HOMO) and
lowest unoccupied (LUMO) levels. However, this charging effect is not
readily observed experimentally, because it does not affect the I-V
directly. Although charging makes the potential profile asymmetric, the
I-V is symmetric with drain bias because the sense of the asymmetry
reverses perfectly on reversing bias, so that the I-V does not show any
sign of rectification \cite{rMujica2}.

It is interesting to see how the screening effects get diluted as the
thick wire is progressively taken apart into its constituents.
Fig.~\ref{f8} (left) shows the hexagonal sheath of wires (cross-section
shown in the inset) undergoing such a decimation process.  First we
lose the quantum couplings between adjacent wires, valid at a small
increase in separation $\cal{D}$ between them. This process in essence
localizes the charges to the individual wires, disallowing any
transverse charge rearrangement and making transverse screening quite
inefficient. The corresponding electrostatic potential profile starts
to look unscreened, like a ramp, being affected only by limited
transverse flow of charge within each wire, around the individual
atomic cores.  The coupling at this stage is electrostatic, given by
the inter-wire capacitances, which decrease linearly with $\cal{D}$.
Finally, when ${\cal{D}} \approx L/2$, where $L$ is the wire length,
the capacitive couplings with the source and drain electrodes dominate
over the inter-wire capacitances, and the wire charges image on the
nearer electrodes instead of on neighboring wires. At this stage, the
assembly behaves in essence like unscreened wires, with each of their
individual potential profiles (Fig.~\ref{f8}, right) resembling the
ramp-like profile of an isolated wire (blue dotted line).

\begin{figure}
\vspace{2.4in}
{\includegraphics{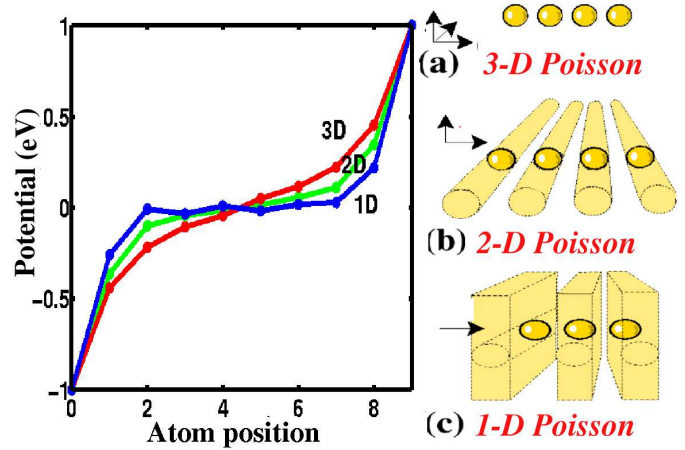}}
\hskip -1.5in
\caption{Dependence of the electrostatic potential on
the dimensionality
of the electric field: (a) single atomic wire simulated with 3-D Poisson's
equation; (b) 2-D distribution of charges and fields, with no transverse
variations, corresponding to a single SAM layer. Each wire atomic charge
is smeared out into a line along the layer direction; (c) variation
allowed only in 1-D, corresponding to a close-packed SAM or a molecular
solid, with the charge on each atom smeared out into a 2-D sheet
along the SAM cross-section. A SAM screens fields in the transverse
direction better,
hence the trend in the potential profile with reducing dimensionality
of the field lines.}
\label{f5}
\end{figure}

In this section, we saw that {\it{screening is ineffective for a
wire that is thinner than its Debye length}}. This is because of the
feasibility of unscreened fields penetrating along the transverse
direction (Fig.~\ref{f4}(d)) that can lead effectively to a poor
overall screening.  Transverse screening can be improved by reducing
the cross-section of the contacts. This eliminates field lines away
from the contacts, generating a Laplace solution that is itself flat
in between the two contacts. In our calculations, however, specifying
the potential boundary conditions on the thin contact cross-sections
alone is not sufficient to allow the method of moments to converge
readily to a solution, due to incomplete specification of boundary
conditions. Convergence requires fixing the potential at a large
distance, as if on imaginary gate electrodes. Similar conclusions
have been obtained for ideal 1-D wires \cite{rJing}.  It has been shown,
for instance, that long-ranged longitudinal screening is ineffective in
conductors such as metallic nanotubes where the charge transport is in
1-D (along the axial direction of the nanotube surface). Thus doping
the nanotube at an interface leads to an asymptotically nonvanishing
surface charge along the nanotube \cite{rTersoff-Odintsov}.

\section{V. Potential screening by the environment: single molecule
vs. monolayer}

\begin{figure*}
\includegraphics[width=5.5in]{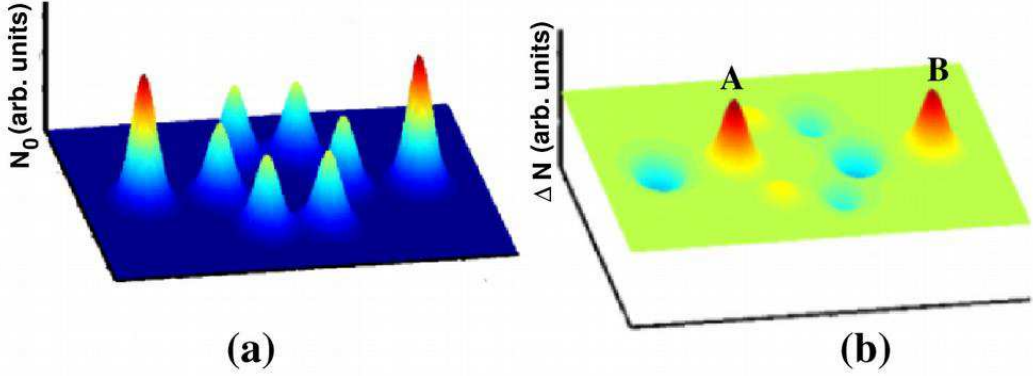}
%\vspace{2.5in}
%{\special{psfile=fig7.ps
%hscale=50.0 vscale=45.0}}
\caption{(a) Electronic charge distribution in PDT at equilibrium,
smoothened artificially using a single Gaussian function for
clarity. There is a distinct barrier on the sulphur atoms due to
electron transfer from gold electrodes. (b)
Change in electron distribution under bias shows an overall electron
flow to the right. Charge piles up at points A and B generating two
residual resistivity dipoles, as current needs to flow around the benzene ring.}
\label{f5_5}
\end{figure*}

\begin{figure}[ht]
\vspace{2.4in}
{\includegraphics{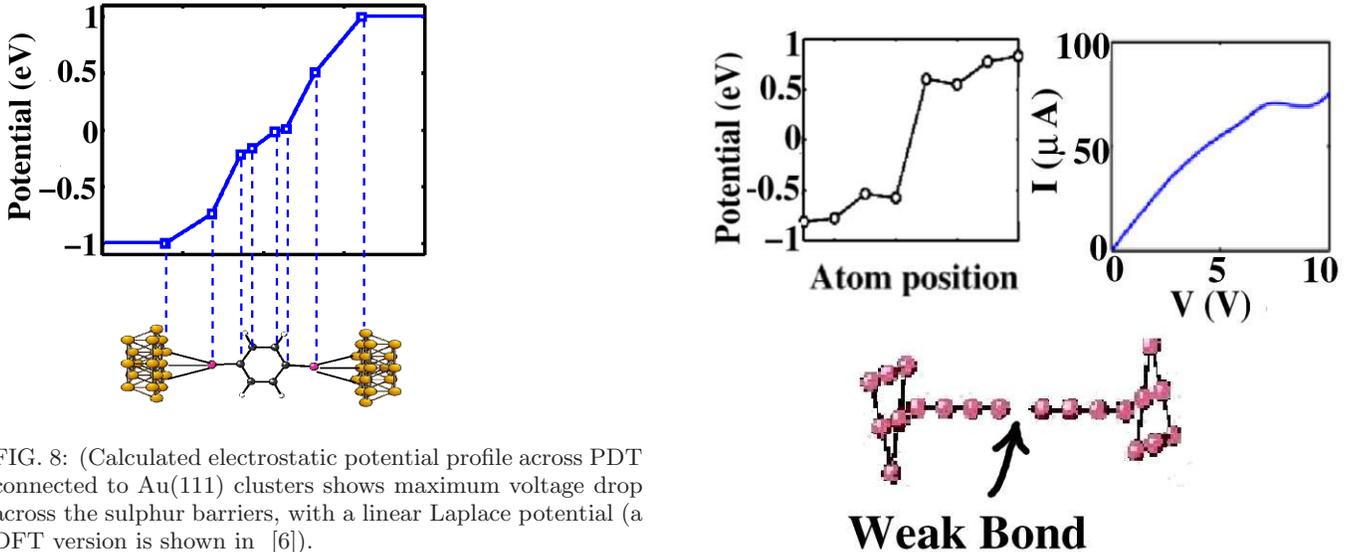}}
\caption{(Calculated electrostatic potential profile
across PDT connected to Au(111) clusters shows maximum voltage drop
across the sulphur barriers, with a linear Laplace potential (a DFT
version is shown in ~\cite{rDamle}).}
\label{f6}
\end{figure}

We have seen that the efficiency of voltage screening depends on the
availability of adequate wire thickness to enable sufficient charge
reorganization in the transverse direction. Thus for thin wires in
isolation, such as in a break-junction experiment, screening is
incomplete and the potential profile is essentially given by the
Laplace part $\phi_L$. In contrast, embedding the wire inside a
self-assembled monolayer (SAM) allows sufficient charge rearrangement
among wires, so that the transverse fields can be efficiently screened
out by neighboring wires.  We investigated transverse screening by
neighboring wires by modeling a SAM that is very densely packed. For
wire separation much smaller than the wire length (half the length to
be precise, as we saw in the previous section), one could ignore charge
and potential variations in the transverse direction.  This process
amounts to replacing the 3-D Poisson equation with a 2-D version that
smears each gold atom into a transverse wire generating a mat of gold
atoms, or a 1-D version smearing each gold atom into a sheet generating
a block or SAM of gold atoms (Fig.~\ref{f5}).  Instead of actually
solving Poisson's equation in reduced dimensions for each case, we
replace the PPP 3-D Green's function in Eq.~\ref{eCharge3} with a 2-D
version $U(\vec{r}_i,\vec{r}_j) \propto \ln\left|\vec{r}_i -
\vec{r}_j\right|$ and with the 1-D version $U(x_i,x_j) \propto min(|x_i
- x_j|,L-|x_i-x_j|)$. For transport, we assume that the wires are
separated and do not have strong transverse couplings, so that one does
not obtain a transverse band of energies. Charge cannot flow easily in
the transverse directions by hopping from wire to wire, and one gets
essentially several copies of the same wire along the transverse
directions. This means that instead of using $F_{\rm{1D}}$ and
$F_{\rm{2D}}$ functions along with the 2-D and 1-D Poisson equations
like we did in section IV, we use $f(E)N_l$ and $f(E)N_A$, where $N_l$
and $N_A$ respectively denote the number of wires per unit length and
the number per unit area.

The resulting self-consistent potentials are shown in Fig.~\ref{f5}.
The inter-wire separations for the 2-D and 3-D cases are assumed to be
equal to the separation between gold atoms (2.885~\AA) within the wire.
From the figure, we see that as the Poisson equation is shifted to
lower dimensions corresponding to the inclusion of more and more
neighboring wires, the potential profile gets progressively flatter due
to enhanced screening efficiency of the transverse modes, so that the
1-D Poisson profile appears well-screened \cite{rRatner}.

\section{VI. Intramolecular potential variations}

The previous sections dealt with the broad shape of the potential
profile across a molecular conductor. As was evident from Fig.~\ref{f4},
the qualitative features of the potential profile can basically be
understood in terms of simple semiclassical continuum pictures, except
for Friedel oscillations and issues relating to the molecular density
of states and single-electron charging. The question may arise at this
stage as to whether the sophisticated machinery of quantum transport
and NEGF, or the quantum Hamiltonian and atomicity of the device are
important at all and whether they could have observable consequences
beyond what is predicted from a continuum description. The answer is that
it could matter, specifically for processes that involve the chemistry
of the molecule.  Specifically, we will discuss two such examples: (i)
potential barriers in PDT, and (ii) features related to the alignment
of levels localized on different parts of the molecule, such as those
generating a diode-like I-V or a negative differential resistance (NDR).

\begin{figure}
\vspace{3.2in}
{\includegraphics{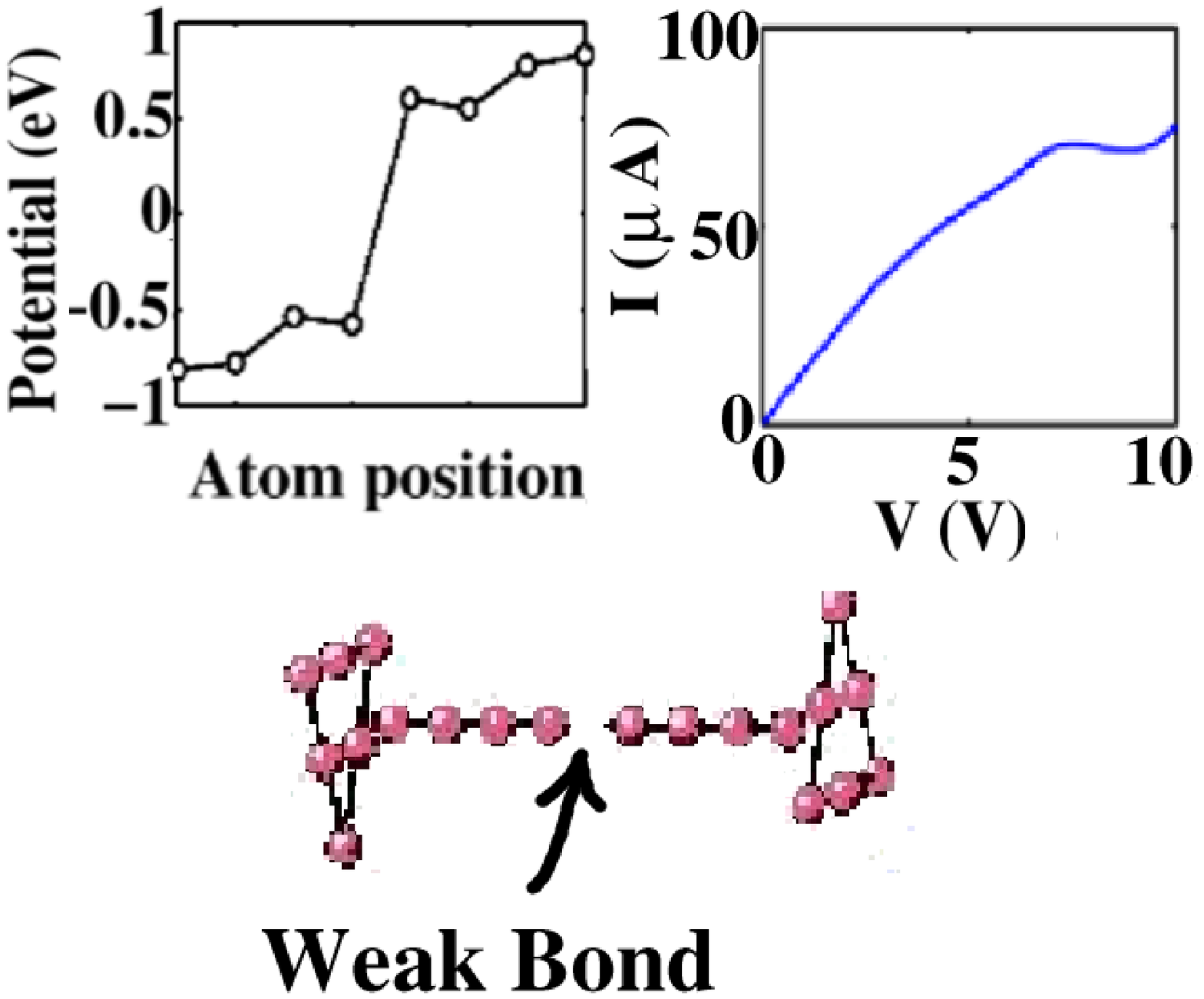}}
\caption{Most of the applied bias drops across a weakened bond in a QPC
(top left), effectively separating the two sides which individually
equilibrate with the corresponding metal contacts. The QPC I-V shows a
weak NDR (top right) due to voltage-induced 
alignment and misalignment of the LDOS on
both sides of the defect (a DFT version is shown in \cite{rNDR}).}
\label{f6.5}
\end{figure}

Fig.~\ref{f5_5} shows the electronic charge distribution on PDT
at equilibrium and under a 2 volt bias.  Fig.~\ref{f6} shows the
corresponding electrostatic potential profile, including gold clusters at
the end. The result, obtained from a semiempirical calculation, compares
well with DFT calculations performed elsewhere \cite{rDamle}.  The largest
voltage drop occurs between the end sulphur and gold atoms, with the
sulphur atoms acting as a barrier (Fig.~\ref{f5_5} and Fig.~\ref{f6}),
and minimal voltage drop within the gold cluster itself. The sulphur
barrier is formed because of the ionic Au-S bonds, which involve
transfer of electrons from the gold leads onto the electronegative
sulphur atoms. Although it is hard to measure such an atomic barrier
experimentally, it is instructive to note that the barriers on sulphur
do control the magnitude of the current to some extent, although the
thinness of the barrier (few Angstroms) probably allows substantial
electron tunneling through it. Furthermore, the barrier regulates the
flow of current, causing charge to pile up there under bias forming a
residual resistivity dipole, as the current rearranges to go around the
benzene ring (Fig.~\ref{f5_5}).

Atomistic barriers and defects can regulate the directionality and
magnitude of current flow patterns within the molecule. In addition,
coherent transport of electrons leads to Friedel oscillations in the
potential profile around the defects and interfaces (Fig.\ref{f4}b;
also see \cite{rDamle,rRRatner}).  This is a purely quantum mechanical
phenomenon arising from the sharpness of the Fermi surface at low
temperature. The Friedel oscillations are superposed on the
long-wavelength Debye screening described earlier. The coherent
oscillations can be eliminated by incorporating incoherent scattering
into the molecule through B\"uttiker probes (\cite{rDamle,rRRatner}).

The intramolecular electrostatic potential determines the local energy
levels and the local density of states (LDOS), so that any substantial
potential variation could bring various parts of the molecule in and
out of resonance. Such resonances can lead to observable consequences,
such as generating an asymmetric, diode-like I-V by bringing the donor
and acceptor levels at two ends of a molecule in and out of resonance
(Aviram-Ratner mechanism) \cite{rAviramRatner}. An analogous process
is shown in Fig.~\ref{f6.5}, where an artificially weakened bond in a
QPC leads to maximum potential drop across it, separating the wire into
two parts that separately equilibrate with the gold leads they are in
contact with. The LDOS of the two parts slide past each other due to
$\phi_L(\vec{r})$, bringing transmission peaks on both sides in and
out of resonance. Such a resonance leads to a weak NDR \cite{rpauls}
(a DFT-based version was demonstrated earlier, in \cite{rRRatner}).

\section{VII. Influence of remote contacts: gate modulation}

\begin{figure*}
\vspace{5.5in}
{\includegraphics{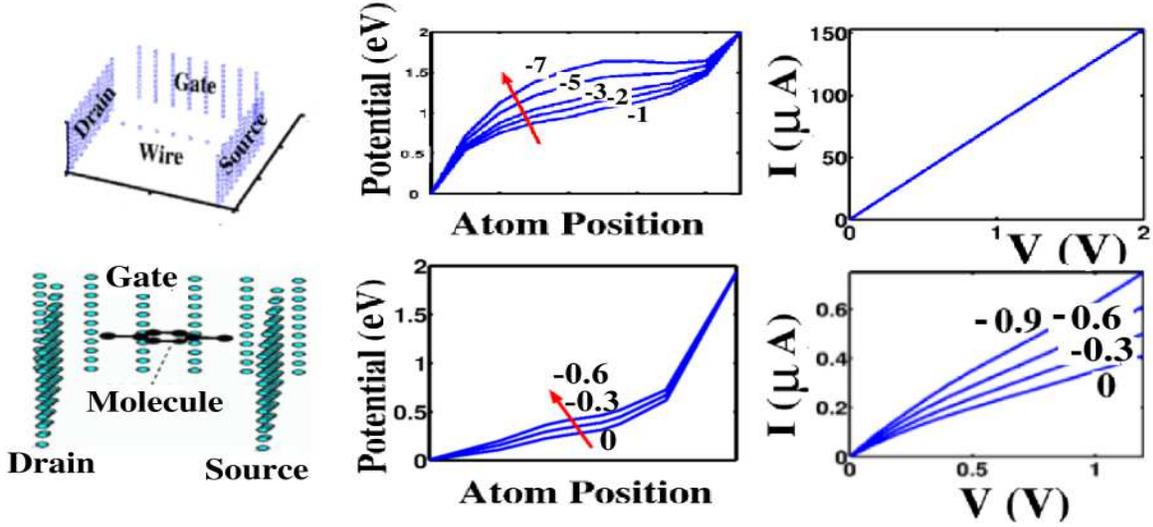}}
\hskip -5.5in
\vskip -3.0in
\caption{Influence of a gate on the molecular I-V for (a) a gold wire
and (b) a PDT molecule placed near the gate (gate is placed
1.5 nm away from the gold wire and 2.9 $\AA$~from PDT). The gate controls
the device potential and shifts the molecular levels relative
to the contact electrochemical potentials. The current is
unaffected for a gold wire due to its relatively featureless
density of states that make it insensitive to level shifting.
For PDT, the current tends to saturate when the drain
electrochemical potential ventures into the HOMO-LUMO gap. The
gate control, however, is quite poor due to the poor aspect-ratio
in the problem, as well as metal-induced gap states (MIGS) from
the gold contacts. Furthermore, the system is expected to have
considerable gate leakage due to the thinness of the gate insulator
(vacuum in this treatment), making the molecule a poor transistor.
The gating can be improved in principle by using longer molecules
with doped silicon contacts as source and drain, and high-k gate
insulators to eliminate gate leakage currents.}
\label{f3} 
\end{figure*}

We have seen that the conductance of a molecule can be influenced
simply by the Laplace part of the electrostatic potential, which
depends on the geometry and shape of the source and drain contacts. In
addition, the potential profile in a conductor can be substantially
influenced by the presence of a third (gate) electrode (Fig.~\ref{f3},
top left). Image charges on the gate electrode deemphasize the Poisson
solution, so that the gate-induced molecular potential is essentially a
Laplace contribution $\phi_L$. A negative gate bias raises the
molecular energy levels by increasing the average electrostatic
potential $\phi_L$ (Fig.~\ref{f3}, top center) relative to the contact
Fermi energy. For a molecule with constant DOS such as a thin gold
wire, this shift does not affect the I-V, which remains ohmic with a
quantized conductance $G_0 = 2e^2/h \approx 77 \mu$S for varying gate
voltage values (Fig.~\ref{f3}, top right). The relative insensitivity
of the conductance quantization to the contact geometry has been
discussed at length in several papers \cite{rBeen}. However, for a
transistor involving a molecule such as PDT (Fig.~\ref{f3}, bottom
left) having a DOS with a lot of structure associated with broadened
HOMO and LUMO levels, the overall potential shift affects the zero-bias
conductance. For PDT, the conductance increases with increasing
negative gate voltage because the molecule is essentially p-type
(closest conducting level to $E_F$ is HOMO-based). In addition, the
potential profile is skewed towards the source-end (Fig.~\ref{f3},
bottom center), which has a common ground with the gate electrode. 
Good gate control in a ``well-tempered'' metal oxide semiconductor field
effect transistor (MOSFET)
makes the Laplace potential insensitive to
source-drain bias and sets it by the source-gate bias
instead (as in Fig.~\ref{f3}) \cite{rlund}. The asymmetric potential profile
effectively tends to pin the channel potential to the source
electrochemical potential, so that only the electrochemical potential
of the drain electrode varies under source-drain bias.  For negative
drain bias, the drain electrochemical potential $\mu_D$ enters the molecular
HOMO-LUMO gap (HLG), leading to an I-V characteristic with decreasing
slope due to a decreasing DOS in the gap (Fig.~\ref{f3}, bottom right).
Different gate voltages yield different zero-bias values $E_F$ for the
drain electrochemical potential $\mu_D$ relative to the levels. This gives a
gate-voltage dependence of the saturation current, leading to MOSFET-like
I-V characteristics with gate modulation \cite{rDamle2,rlangemb}.

While the above gate control mechanism describes the principle of
operation of an ideal ballistic silicon MOSFET \cite{rlund}, the
quantitative conclusions (saturation, gate modulation) are usually
severely compromised when the MOSFET is scaled to molecular
dimensions.  Electrostatic gate control requires
the gate electrode to be substantially closer to the molecule than the
source-drain electrodes, while good saturation in the IVs (large
impedances) require a vanishing DOS in the HLG.  For long molecules
\cite{rfrisbie} one can have a modest oxide thicknesses that still
yields appreciable gate control. However, for small molecules $\sim 10$
\AA, the oxide thickness cannot be scaled down enough without
causing dielectric breakdown, leading to poor electrostatic gate
control \cite{rDamle2}. Furthermore, the current saturation is poor
(Fig.~\ref{f3}) due to a non-negligible gap DOS arising from the broad
HOMO tails generated by MIGS from the gold contacts. Improving the gate
control and the impedance require the utilization of high-k gate
dielectrics and degenerately doped semiconducting contacts.
Alternately, one could envision transistor action based on
non-electrostatic principles, such as by gating the molecular
conformations, for example \cite{rTitAvi}.

The gate control can be enhanced by wrapping the molecule with
cylindrical gates surrounding the wire.  Such a scheme is employed in
silicon transistors to produce dual-gate MOSFETS \cite{rHu0}, FINFETs
\cite{rHu1} or structures with wrap-around gates \cite{rHu2}.  We 
have modeled
the effect of multiple gates by placing two, three and four rectangular
gate electrodes symmetrically around the wire. Increasing the number of
gate electrodes increases the potential barrier, along with an overall
increase in the average potential. This leads to superior gate control.
The practicality of this scheme, however, depends on the capacity to
have multiple gates closely spaced around the molecular wire and 
insulated from it.

\section{IX. Conclusions}
The electrostatic potential profile across a conductor controls the
distribution and magnitude of current across it. While broader features such
as the overall conductivity and screening lengths determine the shape
of the I-V curves in many cases, atomistic features involving level
resonances can influence these I-Vs in a nontrivial way, necessitating a
proper atomistic, quantum kinetic treatment of charge transport in
molecular devices. We have performed such a treatment within a coupled
Poisson-NEGF formalism, while simplifying the molecular chemistry to
essentially one orbital per atom for computational simplicity and ease
of illustration (a more sophisticated DFT version for some of these
profiles has been demonstrated elsewhere).  We have seen that the
Poisson part of the electrostatic potential profile depends on the
strength of transverse and longitudinal screening, and can yield
nontrivial charging effects. A gate electrode generates image charges
that deemphasize the role of the Poisson part, while modulating the
Laplace solution suitably to significantly alter the device I-V
characteristic and yield saturating currents. However, feasibility of
electrostatic gate control depends on the channel length to oxide
thickness aspect-ratio in the problem, and may need to be supplanted by
alternate physical principles, such as employing conformational degrees
of freedom for example. 

We would like to thank P. Damle, J. Guo, T. Rakshit, R. Venugopal,
A. Nitzan, P. Solomon and F. Zahid for useful discussions. This work has been
supported by ARO-DURINT, Award\# 527826-02, and NSF, Award\# 0085516-EEC.


\begin{thebibliography}{100}
\bibitem{rDebye} For a 3-D noninteracting
electron gas this gives $\lambda_D = \sqrt{2\epsilon E_F/3n_{\rm{eq}}e^2}$, where $E_F$ is
the Fermi energy and $n_{\rm{eq}}$ is the equilibrium electron density.
For a classical electron gas at temperature $\cal{T}$ one simply replaces 
$E_F \rightarrow 3/2k_B\cal{T}$.
\bibitem{rpn} A classic example is a p-n junction at equilibrium, where the
electrochemical potential is constant while the electrostatic potential varies
over a depletion width, generating a built-in electric field across the junction. 
Note that for a semiconductor, the conductivity $\sigma$ depends on the potential
$\phi$ through band-bending.
\bibitem{rGauge} The choice of values for $\mu_{S,D}$ is
gauge-invariant. If we add a constant to the contact electrochemical
potentials, the extra charge on the contacts would cause the contact
levels to float up, so that the added constant will get incorporated
into the Laplace part $\phi_L$ of the electrostatic potential.
\bibitem{rnote} The evolution of the exchange-correlation part can be
taken into account if instead of just iterating the Poisson's equation,
one iterates a proper DFT Hamiltonian with appropriate boundary
conditions, self-consistently coupled with NEGF. Such a scheme has
been used in \cite{rDamle}, with an improved treatment of the boundary
conditions for the Hartree part described in \cite{rDamle3}. Here,
we concentrate on just the evolution of the Hartree part under bias,
in keeping with the main focus of our paper.
\bibitem{rab} The dimensional subscript for the $F$ functions describes
the trivially summed dimensions, and when combined with the dimension
of the explicitly solved Poisson's equation, should add up to three.
\bibitem{rcontregions} As explained in section III, including built-up regions
of the contact within the device leads automatically to a geometric dilution
that reduces the amount of the applied voltage lost within the contacts. This
is confirmed by increasing the size of the contact region, which does not 
affect the voltage dropped across the wire. The potential profile inside the
contact flattens out and reaches the applied bias within a couple of layers
from the wire surface.
\bibitem{rlangemb} This mechanism of current saturation and gate modulation
is the analogue of the corresponding mechanism in a ballistic MOSFET \cite{rlund}, 
and dominates
over other second-order effects such as charging of the surface atoms 
\cite{remberlyfet} or simple Stark shifts of the molecular levels \cite{rlangfet}.
\end{thebibliography}
\end{document}